\documentclass[manuscript]{aastex}

\slugcomment{The Astronomical Journal}

\shorttitle{}
\shortauthors{Kasuga and Jewitt et al.}

\begin{document}


\title{Observations of 1999 YC and \\
the Breakup of the Geminid Stream Parent}


\author{Toshihiro Kasuga\altaffilmark{1,2} and 	
	David Jewitt \altaffilmark{1}}
\email{kasugats@ifa.hawaii.edu, jewitt@ifa.hawaii.edu}





\altaffiltext{1}{Institute for Astronomy, University of Hawaii,
               2680 Woodlawn Drive, Honolulu, HI 96822}
\altaffiltext{2}{JSPS Research Fellow, National Astronomical Observatory
              of the Japan, National Institute of Natural Science,
              2--21--1 Osawa, Mitaka, Tokyo 181--8588, Japan}



\begin{abstract}
Apollo asteroid 1999 YC may share a dynamical association with the
Phaethon-Geminid stream complex \citep{Ohtsuka2008}.  Here, we present
photometric observations taken to determine the physical properties of
1999 YC.  The object shows a nearly neutral reflection spectrum, similar
to but slightly redder than related objects 3200 Phaethon and 2005 UD.
Assuming an albedo equal to 3200 Phaethon we find that the diameter is
1.4$\pm$0.1\,km.  Time-resolved broad-band photometry yields a
double-peaked rotational period of 4.4950\,hr $\pm$0.0010 while the
range of the lightcurve indicates an elongated shape having a projected
axis ratio of $\sim$ 1.9.  Surface brightness models provide no evidence
of lasting mass loss of the kind seen in active short period cometary
nuclei.  An upper limit to the mass loss is set at
$\sim$$10^{-3}$$\,{\rm kg\,s^{-1}}$, corresponding to an upper limit on
the fraction of the surface that could be sublimating water ice of $ < $
$10^{-3}$.  If sustained over the 1000 yr age of the Geminid stream, the
total mass loss from 1999 YC (3$\times$10$^7$ kg) would be small
compared to the reported stream mass ($\sim 10^{12}$ -- $10^{13}$\,kg),
suggesting that the stream is the product of catastrophic, rather than
steady-state, breakup of the parent object.
\end{abstract}


\keywords{comets: general --- minor planets, asteroids --- meteors: general}



\section{Introduction}


Breakup or disintegration is an apparently common end-state for the
nuclei of comets (Chen and Jewitt 1994, Boehnhardt et al. 2004, Jewitt
2004).  Meteoroid streams appear to result from this process, including
some which are associated with parent bodies whose physical properties
are those of asteroids, not obviously comets
\citep{baba1994,jenniskens2006}.  A leading example of the latter is the
Apollo type near- Earth asteroid 3200 Phaethon (1983\,TB) which is
dynamically associated with the Geminid meteoroid stream
\citep{Whipple1983}.  Phaethon may be a dead or dormant comet, but its
unusual blue reflection spectrum distinguishes it from most other
well-studied cometary nuclei, and no outgassing or mass-loss activity
has ever been reported \citep{Hsieh2005,Wiegert2008}.  Recently,
\citet{Ohtsuka2006} suggested the existence of a ``Phaethon-Geminid
stream Complex (PGC)'' which implies a big group of split cometary
nuclei/fragments based on their dynamical similarity.  They identified
asteroid 2005 UD, classified as an Apollo-type, as having a common
origin with 3200 Phaethon \citep{Ohtsuka2006}.  Subsequent photometry of
2005 UD revealed that it has unusual blue optical colors like those of
Phaethon, consistent with the idea that these two bodies have a common
origin \citep{jewitt2006,Kinoshita2007}.

Recently, \cite{Ohtsuka2008} suggested that another apparently
asteroidal object, 1999 YC, may have an orbital association with 2005
UD, 3200 Phaethon and the Geminid meteoroid stream.  In this paper we
present physical observations of 1999 YC.  Its colors, size, rotational
period, limits to the on-going mass loss rate and the fractional active
area area compared with those of 2005 UD and 3200 Phaethon.

\section{Observations}
Observations were taken on the nights of UT 2007 Sept 4, Oct. 4, 12, 18
and 19 using the University of Hawaii 88-inch (2.2\,m) diameter
telescope (hereafter, UH\,2.2) and the Keck-I 10\,m diameter telescope,
both located at 4200 m altitude atop Mauna Kea, Hawaii.  The UH\,2.2
employed a Tektronix 2048 $\times$ 2048 pixel charged - couple device
(CCD) camera at the f/10 Cassegrain focus.  This detector has an image
scale of $0^{''}.219$ ${\rm pixel}^{-1}$ and a field of view of
approximately $7^{'}.5 \times$ 7$^{'}$.5.  On Keck-I, the LRIS (Low
Resolution Imaging Spectrometer) camera ~\citep{Oke1995} was used in its
imaging mode.  This camera was equipped with two separate cameras having
independent CCD imagers.  One is a blue-side CCD having 4096 $\times$
4096 pixels, each $0^{''}.135$ ${\rm pixel}^{-1}$ and the other is a
red-side detector having 2048 $\times$ 2048 pixels, each $0^{''}.215$
${\rm pixel}^{-1}$.  The blue-side of LRIS was used to record images in
the B filter, while images in the V and R filters were recorded using
the red-side detector of LRIS at the same time, which doubles the
observational efficiency relative to a single-channel camera.

All images were obtained through the Johnson-Kron-Cousins BVR-filter
system with the telescopes tracked non-sidereally to follow the motion
of 1999 YC at rates about ${\rm 52^{''}\,hr^{-1}}$.  Unfortunately,
images in the I-filter could not be obtained at UH\,2.2 due to the
faintness of 1999 YC and 2005 UD while the I-filter at the Keck-I 10\,m
broke shortly before our observing time.  Images were corrected by
subtracting a bias image and dividing by a bias-subtracted flat-field
image.  Flat-field images at the UH\,2.2 were constructed from scaled,
dithered images of the twilight sky.  At the Keck-I 10\,m we obtained
flat-fields using an artificial light to illuminate the inside of the
Keck dome.  Photometric calibration was obtained using standard stars
from ~\cite{Landolt1992}, including 94-401, 95-98, 912-410, 912-412,
L98-627, L98-634, L98-642, L98-646, Mark A1, A2, and A3, and PG2213-006A
and C.  The median full width at half-maximum (FWHM) measured on 1999 YC
varied from $\sim$ $0.6''$ -- $0.9''$.  An observational log is given in
Table~\ref{log}.  Photometry of 2005 UD and 3200 Phaethon was taken in
parallel with 1999 YC in order to compare with published observations
and to minimize the possibility of systematic differences.

\section{Observational Results}
\label{bozomath}

\begin{figure*}[htbp]
\epsscale{0.9} \plotone{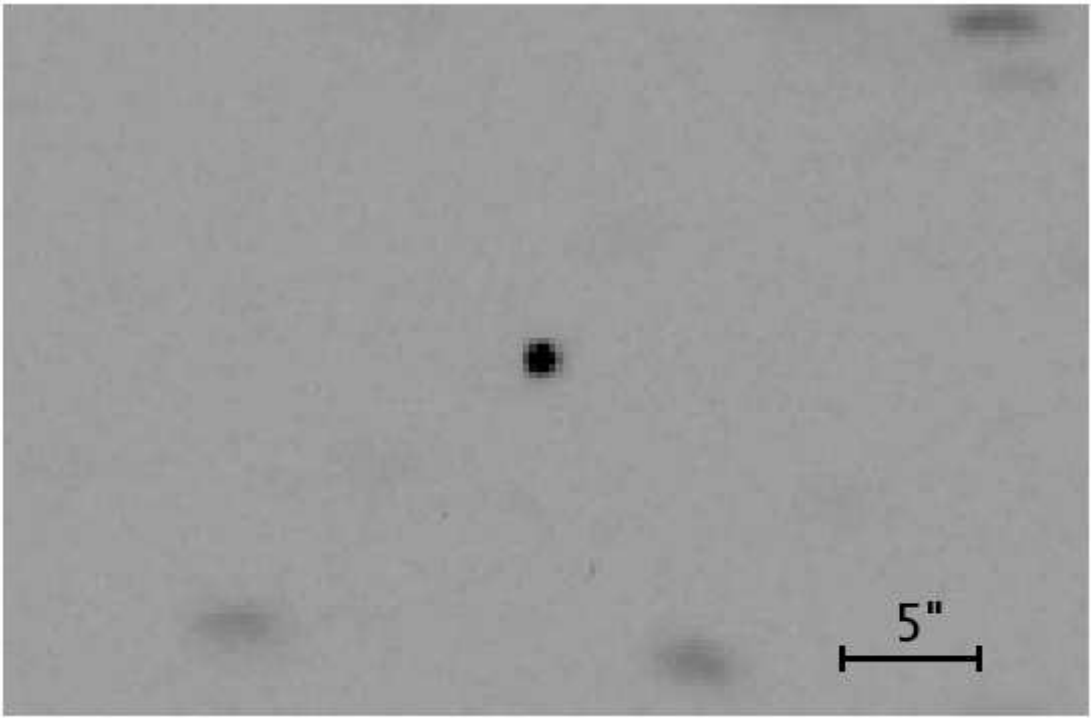} \caption{Median combined ${\it R}$-band
image of the asteroid 1999 YC taken by Keck-I 10\,m on UT 2007 October
12.  The image having a total integration time of 400\,sec shows no
coma.  The object with FWHM of $\sim$ $0.65''$ is centered within the
frame of $40''$ wide.  \label{fig1}}
\end{figure*}

All three objects 1999 YC, 2005 UD and 3200 Phaethon show point-like
images in our data (see Figure \ref{fig1}).  Photometry was performed
using synthetic circular apertures projected onto the sky.  Photometry
was determined using apertures of radius $\sim$ $1.5^{''}$ (typically
two times the image FWHM), while the sky background was determined
within a concentric annulus having projected inner and outer radii of
$3.3^{''}$ and 6.6$^{''}$, respectively.  Photometric results for 1999
YC are listed in Tables \ref{colphot} and \ref{lightphot}. Colors of
1999 YC are given in Table \ref{color} together with those of 2005 UD,
3200 Phaethon and solar color indices.

\begin{figure*}[htbp]
\epsscale{0.9} \plotone{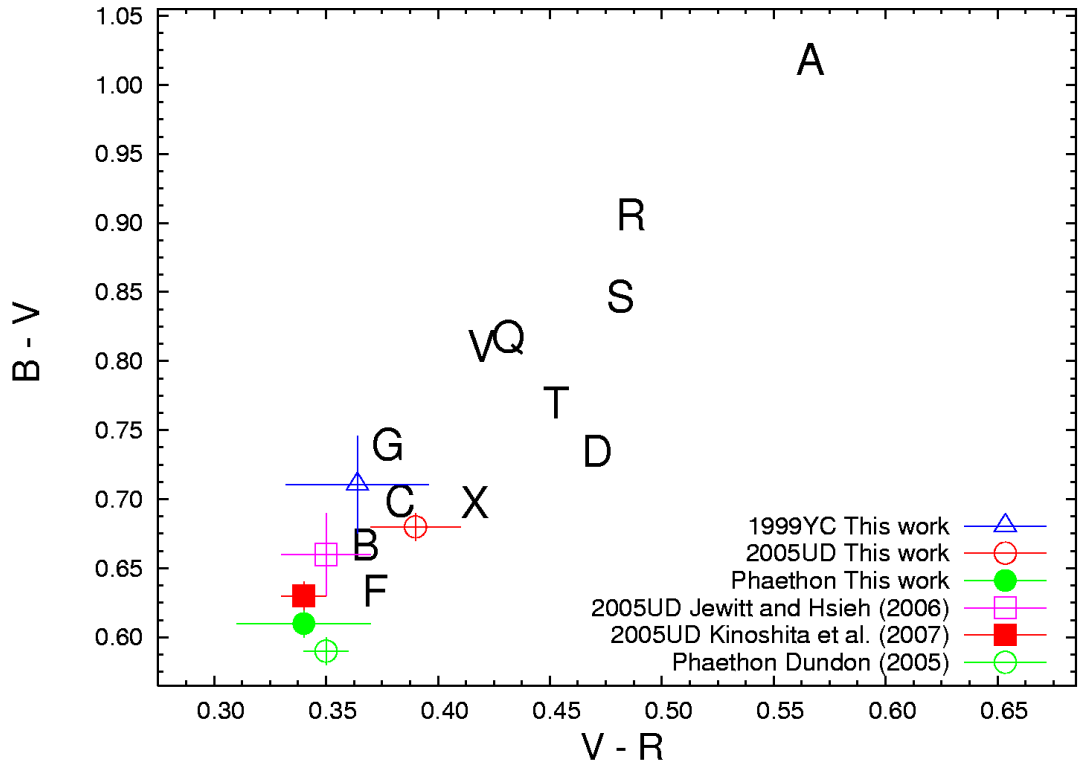} \caption{Color plots of V--R
vs. B--V.  Phaethon-Geminid Complex with the typical trends of near
Earth asteroids qualified as spectrally for Tholen taxonomic classes
\citep{Dandy2003}.\label{test}}
\end{figure*}

Figure \ref{test} shows the relation between B--V and V--R for various
Tholen taxonomy classes measured from 56 near-Earth asteroids
\citep{Dandy2003}.  The colors of 1999 YC are consistent with C-type
asteroids within the uncertainties, and 2005 UD in this work is also
similar to the flattened reflectance spectra as seen in C-type
asteroids.  The colors of 3200 Phaethon measured here agree with
numerous independent studies that spectrally classified it as a B- (or
F-) type asteroid
\citep{Tholen1985,Luu1990,Skiff1996,Lazzarin1996,Hicks1998,Dundon2005,Licandro2007}.

Most published optical colors of 3200 Phaethon and 2005 UD are slightly
bluer than the sun.  Bluer colors are uncommon amongst near-Earth
asteroids, occuring in about 1 of 23 objects \citep{jewitt2006}.
\cite{Kinoshita2007} found that the color of 2005 UD varies slightly
with rotational phase, being bluer than the Sun for 75\% of the
lightcurve but neutral (C-type) for the remainder.  They speculated that
the surface of 2005 UD is heterogeneous perhaps as a result of the
splitting phenomenon or of a collision, consistent with being a PGC
fragment.  Heterogeneity on the surface of 3200 Phaethon was also
suggested \citep{Licandro2007}.  It shows possible spectral variability
due to inhomogeneous compositions caused by the thermal alteration at
its small perihelion distance q$\sim$ 0.14\,AU or by the hydration
process \citep{Licandro2007}.  Therefore, while 1999 YC appears slightly
redder than the rotationally averaged colors of 2005 UD and 3200
Phaethon (Figure \ref{test}), the differences are not much larger than
either the measurement uncertainties or the reported color variations on
2005 UD.  We conclude that the color data are not inconsistent with an
association between 1999 YC and the other objects in the dynamically
defined PGC (see Table \ref{color}).

\subsection{Size}
\label{sizeS}

Table \ref{lightphot} shows the results of R-band photometry of 1999 YC
on the nights of UT 2007 Oct. 4, 18 and 19.  The apparent red magnitude
$m_{\rm R}$ was corrected to the absolute red magnitude ${\rm m}_{\rm
R}$(1,1,0) using
\begin{equation}
m_{\rm R}(1,1,0) = m_{\rm R} - 5\,{\rm log}(r\, \Delta) -\beta \alpha
\label{R}
\end{equation}
where r and $\Delta$ are the heliocentric and geocentric distances in
AU, $\alpha$ is the phase angle in degrees (Sun - Target - Observer) and
$\beta$ is the phase coefficient.

We did not sample the full rotational lightcurve variation on each epoch
of observation.  Therefore, we cannot simply use the mean or median
brightness at each epoch in order to measure the phase variation.
Instead, we determined the linear phase coefficient, $\beta$ in
mag.\,${\rm deg}^{-1}$, from the better-observed brightness maxima in
night-to-night light curves from $m_{\rm R}$ in Table \ref{lightphot}
($N$= 4, 5 on Oct.4, N= 14, 15, 16, 30, 31, 32, 56, 57, 58 on Oct.18,
62, 63, 64 on Oct.19, where $N$ shows the file number in Table
\ref{lightphot}).

\begin{figure*}[h]
\epsscale{0.9} \plotone{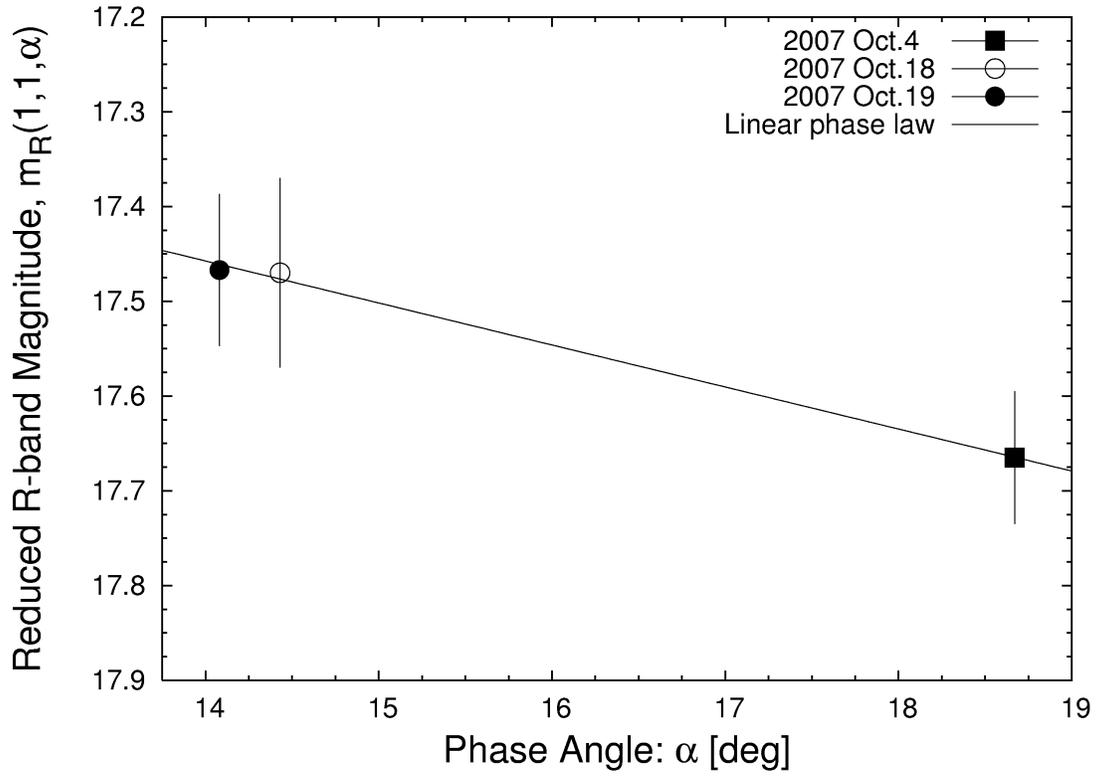} \caption{Linear phase function for
1999 YC.  The solid curve shows a phase coefficient
$\beta$=0.044$\pm$0.002 mag\,${\rm deg}^{-1}$ determined by fitting data
from UT 2007 Oct. 4, 18 and 19.  \label{phase}}
\end{figure*}

Figure \ref{phase} shows the correlation between phase angle and the
reduced apparent red magnitudes corrected to R=$\Delta$=1\,AU by
equation (\ref{R}).  The derived coefficient, $\beta$=0.044$\pm$0.002
mag\,${\rm deg}^{-1}$, is consistent with a low-albedo, as observed in
cometary nuclei \citep{Lamy2004}, and similar to the C-type asteroids
($\beta$=0.041$\pm$0.003) at phase angles of 5$^{\circ}$ $<$ $\alpha$
$<$ 25$^{\circ}$ \citep{Belskaya2000}.

We used the absolute magnitudes, ${\rm m}_{\rm R}$(1,1,0), to calculate
the equivalent circular diameter, $D_{\rm e}$, using
~\citep{Russell1916}

\begin{equation}
D_e [km] = \left[\frac{1140}{p_v^{1/2}}\right] 10^{(-0.2m_R(1,1,0))}
\end{equation}

\noindent in which we have taken the apparent red magnitude of the Sun
as $m_R$ = --27.1 \citep{Cox2000}.  We adopt the albedo $p_v(\approx
p_R)$ = 0.11$\pm$0.02 as obtained from infrared observations of 3200
Phaethon ~\citep{Green1985},

Table \ref{size} lists the absolute magnitudes ${\rm m}_{\rm R}$(1,1,0)
and the resulting equivalent circular diameters $D_{\rm e}$ of 1999 YC,
2005 UD and 3200 Phaethon.  The Table shows that 1999 YC and 2005 UD are
similar in size and each is about one quarter of the diameter (and,
presumably, 4$^{-3}$ $\sim$2\% of the mass) of 3200 Phaethon.

\subsection{Light curve}

\begin{figure}[htbp]
\epsscale{0.9} \plotone{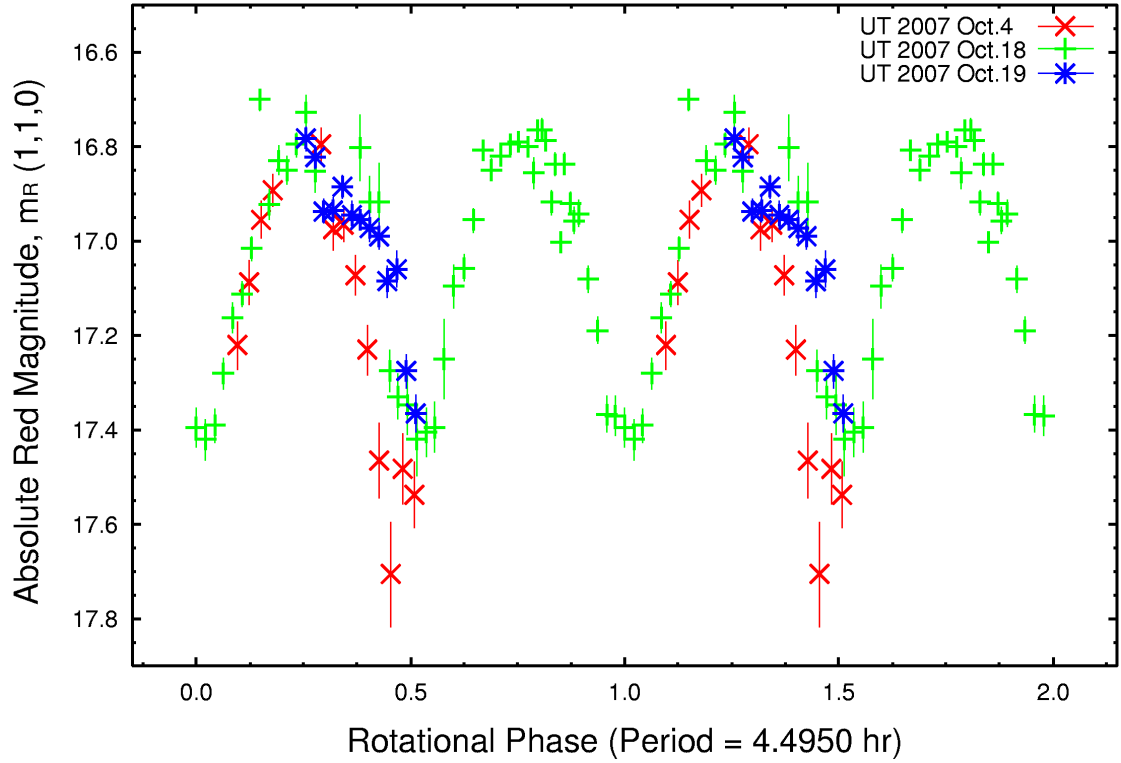}\caption{Rotational phase vs. absolute
 red magnitude variation of 1999 YC observed on UT 2007 Oct. 4, 18 and
 19.  ${\rm m}_{\rm R}(1,1,0)$ is phased to the double-peaked rotational
 period of $P_{\rm rot}$= 4.4950 $\pm$ 0.0010 \,hr.  \label{Abs}}
\end{figure}

\begin{figure}[htbp]
\epsscale{0.9} \plotone{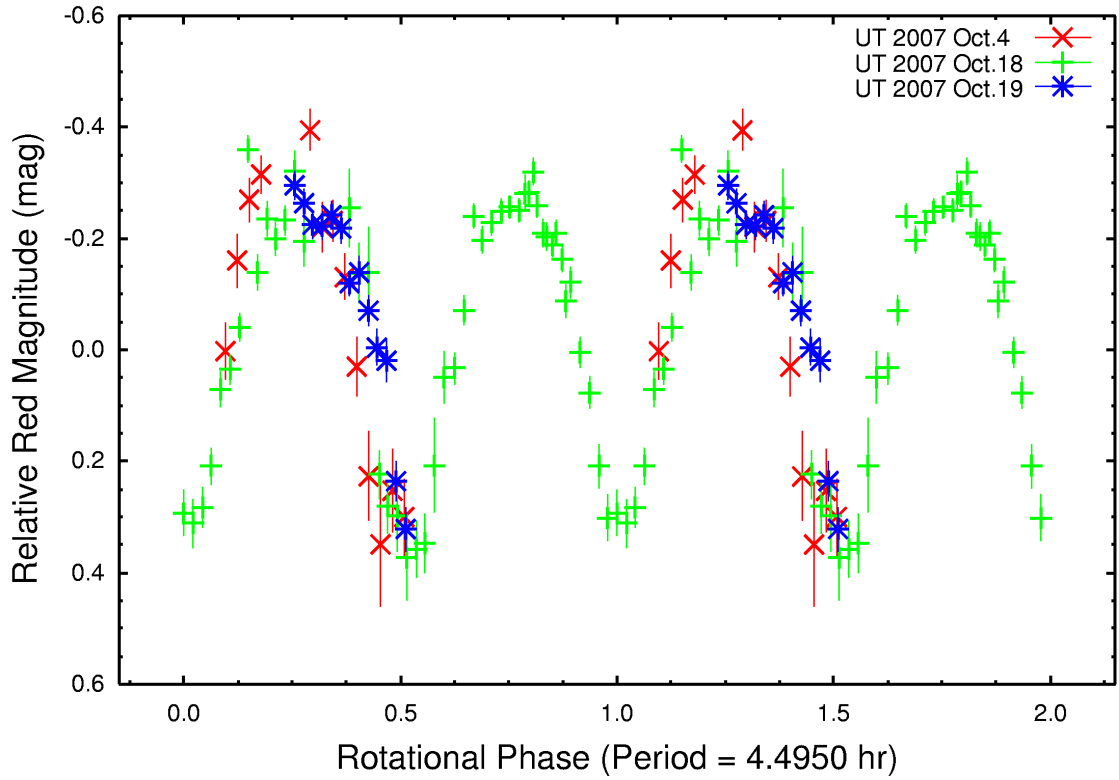} \caption{Rotational phase vs. relative
 red magnitude variation of 1999 YC observed on UT 2007 Oct. 4, 18 and
 19, phased to double-peaked rotational period of $P_{\rm rot}$= 4.4950
 $\pm$ 0.0010\,hr.  Magnitudes from the different nights are phased and
 plotted relative the median magnitude each night. \label{Rel}}
\end{figure}

In order to find the rotation period for 1999 YC, the phase dispersion
minimization (PDM) technique ~\citep{Stellingwerf1978} was used both on
the absolute red magnitudes; ${\rm m}_{\rm R}(1,1,0)$, and on the
relative red magnitudes defined as excursions from the median magnitude
measured each night.  The PDM in the NOAO IRAF software package provided
us with several possible light curve periods and enabled us to visually
examine the data for each period.  The most likely rotational period was
determined by the smallest value of theta (see detail Stellingwerf
1978), namely a single-peaked light curve of period $P_0$=2.247\,hr.
Other possible rotational periods are related to multiples of $P_0$.
The lightcurves of most small bodies in the Solar system are
double-peaked, resulting from elongated shapes rather than from strong
albedo markings.  We assume that the lightcurve of 1999 YC is double-
peaked and so conclude that the true rotation period is $P_{\rm
rot}$=$2P_0$=4.4950\,hr (Figs.~\ref{Abs},~\ref{Rel}).  The uncertainty
on the period is estimated 0.0010\,hr, by examination of the acceptable
phased lightcurves.  This is comparable to the 5.249 hr period of 2005
UD and the 3.59 hr period of 3200 Phaethon (see Table 6).
\citet{Pravec2002} computed the mean spin rate vs. diameter to find
$P_{\rm rot}$ $\sim$ 5.0 $\pm$0.6\,hr for diameters $D_e$ $\sim$ 1\,km.
We conclude that, although short, the rotation period of 1999 YC is not
unusual for asteroids of comparable size.

The maximum photometric range of 1999 YC is $\Delta m_R$=0.69$\pm$0.05,
giving a minimum axis ratio of the body.  Assuming that the amplitude is
shown by the largest and smallest faces presented during the rotation of
an elongated body, the ratio of the long to short axis of 1999 YC
projected on the plane of the sky is expressed by

\begin{equation}
10^{0.4 \Delta m_{R}} = \frac{a}{b} = 1.89 \pm 0.09
\end{equation}
where {\it a} is the long axis and {\it b} is the short axis.  While
1999 YC is more elongated than either 2005 UD ({\it a/b}=1.45$\pm$0.06)
or 3200 Phaethon {\it a/b}$\sim$1.45 \citep{jewitt2006, Dundon2005}, the
observed differences cannot be accurately interpreted because of the
unknown spin-vectors of these bodies and the effects of projection.
 
A critical density $\rho_c $ can be derived from $\rho_c $ = 1000 ${\rm
(3.3\,hr/ P_{rot})^2}$ for a spherical body with a given rotation period
in hour $P_{\rm rot}$, by equating the acceleration of gravity at the
surface with the centripetal acceleration at the equator.  For an
elongated body like 1999 YC, the acceleration of gravity at the tip of
the long axis {\it a} is reduced by a factor about equal to the axis
ratio, {\it b/a}, compared to that of a sphere of the same density and
radius ~\citep{Harris1996,Pravec2000}.  Therefore, a critical density
for an elongated body is described as

\begin{equation}
\rho_c \approx 1000\left(\frac{3.3\,{\rm hr}}{P_{\rm
		  rot}}\right)^2\left(\frac{a}{b}\right). 
\end{equation}

where $P_{\rm rot}$ is in hours.  The critical density is a lower limit
in the sense that a less dense body with the observed period and axis
ratio would be in a state of internal tension against centripetal
acceleration.  The critical densities are compared in Table 6.

One additional feature of the lightcurve shown in Figures \ref{Abs} and
\ref{Rel} is worthy of note.  The data from UT 2007 Oct 04 do not fit
those from UT 2007 Oct 18 and 19 quite so well as the latter two nights
considered alone.  This could simply be because 1999 YC was nearly a
magnitude fainter on the first date of observation, and therefore the
effects of photometric uncertainties in the measurements are
proportionally larger.  Alternatively, it is possible that the slightly
discrepant shape of the lightcurve from UT 2007 Oct 04 is a result of
non-principal axis rotation of 1999 YC.  The latter is to be expected if
1999 YC is a recently produced fragment, because splitting of the
nucleus should naturally produce excited rotational states and the
timescale for the damping of nutation by internal friction is very long
for bodies as small as 1999 YC \citep{Burns1973}.  For example, for a
rubble pile structure, the damping time is $T_{\rm d}$= 0.24 $P_{\rm
rot}$$^3$/$r_{\rm obj}$$^{2}$ (in millions of years) \citep{sharma2005}.
Substituting $P_{\rm rot}$ and $r_{\rm obj}$ for 1999 YC, we estimate
$T_{\rm d} \sim$10$^7$\,yr.  This is longer than the dynamical lifetime
of $10^{6}$\,yr \citep{Froeschle1995} and \textit{much} longer than the
estimated $\sim$$10^{3}$ yr age of the Geminid stream, implying that
precessional motions aquired at formation would not yet have been damped
by internal friction.  The same considerations apply to 2005 UD and 3200
Phaethon and non-principal axis rotation should be a general feature of
these and other PGC fragments.  Still, better temporal coverage will be
needed to unambiguously detect non-principal axis rotations.

\subsection{Surface Brightness model}
\label{SBS}

To search for coma in 1999 YC we compared its measured surface
brightness profile with the profiles of unresolved field stars and with
a seeing-convolved profile of a model comet.  Because of the
non-sidereal motion of 1999 YC, the images of background sources appear
trailed in the data and so the surface brightness must be treated using
the procedures of Luu and Jewitt 1992.

\begin{figure*}[htbp]
\center \epsscale{0.9} \plotone{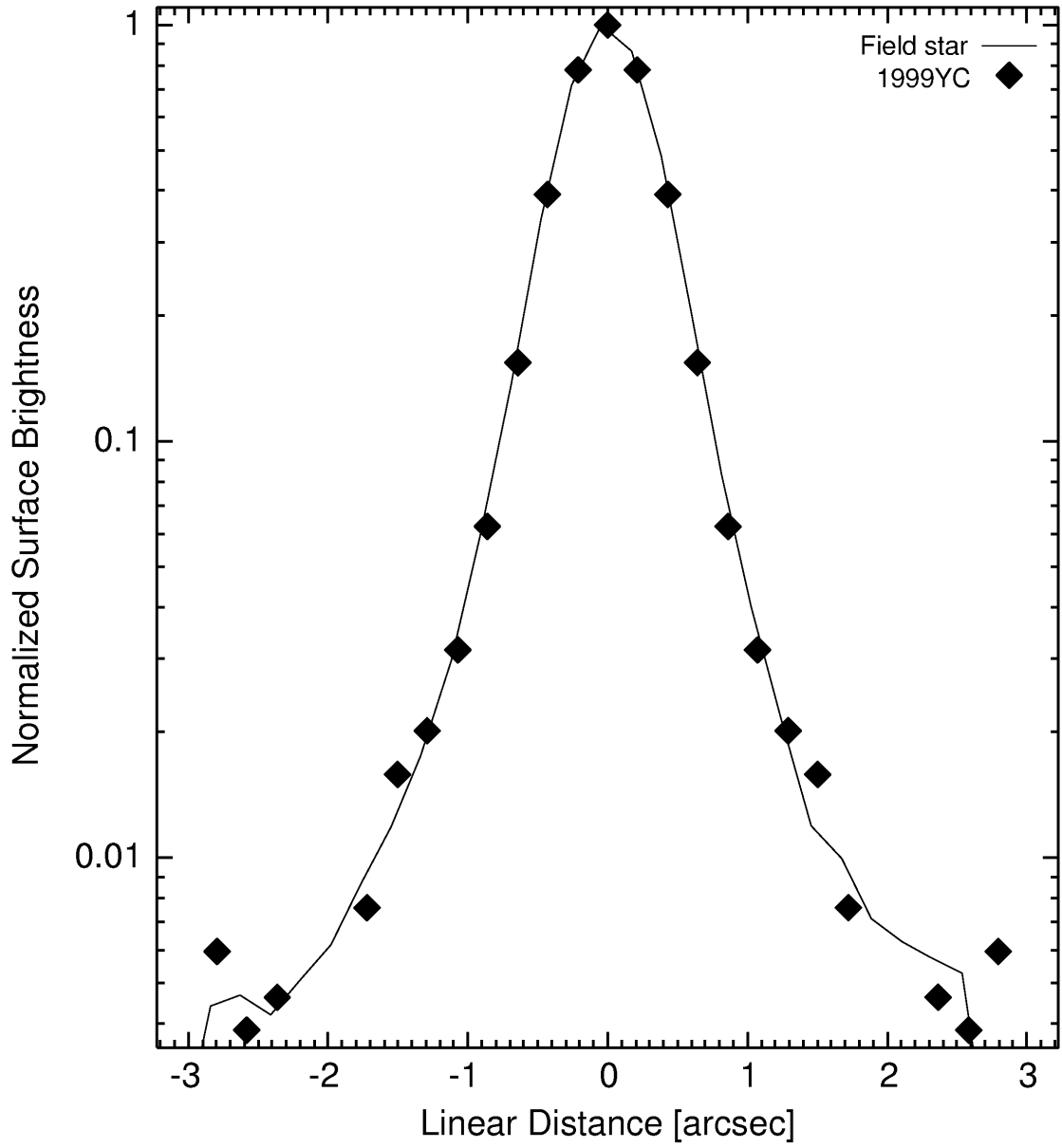}\caption{Normalized 1D surface
 brightness profiles of 1999 YC and a field star. \label{1D}}
\end{figure*}

Firstly, to determine one - dimensional\,(1D) surface brightness
profiles of the asteroid and the field star, we selected two {\it
R}-band images taken using Keck-I on the night of UT 2007 October 12
(combined integration time = 400\,sec, see Fig \ref{fig1}) because the
Keck signal to noise ratio (S/N $\geq$ 100) is greater than from the
UH\,2.2 (S/N $\simeq$ $20 \sim 30$).  Each image was rotated using a
fifth order polynomial interpolation so the field star trail was aligned
parallel to the pixel rows in the image frame, then median-combined into
a single image (having FWHM $\sim$ $0.65''$).  Then, 1D surface
brightness profiles of 1999 YC and a field star were measured in the
direction perpendicular to the trail.  Each profile was averaged along
the rows over the entire width of the asteroid and the field star after
subtracting sky.  Both normalized profiles are similar, although with
small differences attributed to noise in the data (Fig. \ref{1D}).

Secondly, to set quantitative limits to coma in 1999 YC, we compared the
two - dimensional (2D) point spread function (PSF) of the asteroid with
seeing convolved model profiles of the comet.  The seeing was determined
from the 2D PSF of a field star, and convolved with simple comet models
of ``nucleus plus coma''.  In model images of 100$\times$100 pixels, the
nucleus was represented by a spike located at the central pixel and the
spherically symmetric coma.  The parameter $\eta$($\ge 0$) defined as
the ratio of the coma cross section ${\it C_c}$ to the nucleus cross
section ${\it C_n}$, which corresponds to the ratio of the flux density
scattered by the coma ${\it I_c}$ to the flux density scattered by the
nucleus cross section ${\it I_n}$ was able to characterize varying coma
- activity levels on preconvolution models ~\citep{Luu1992}, and
expressed as

\begin{equation}
\eta=\frac{C_c}{C_n}=\frac{I_c}{I_n},\, \eta \ge 0
\label{eta}
\end{equation}

The intensity ${\it I_c}$ of each pixel in the coma was determined by
$I_c=\int_0^{\phi} 2\pi r \cdot K/r \cdot \,dr$ as a function of the
surface brightness, where ${\it K}$ was a constant of proportionality,
${\it r}$ was distance from the nucleus and $\phi$ is the reference
photometry aperture radius of 50 pixels $(10.95'')$.  The parameter
$\eta$ can take $\eta \ge 0$, with $\eta$=0 indicating a bare nucleus
(no coma) and $\eta$ =1 indicating coma and nucleus having equal cross
sections.

\begin{figure*}[htbp]
\center \epsscale{0.9} \plotone{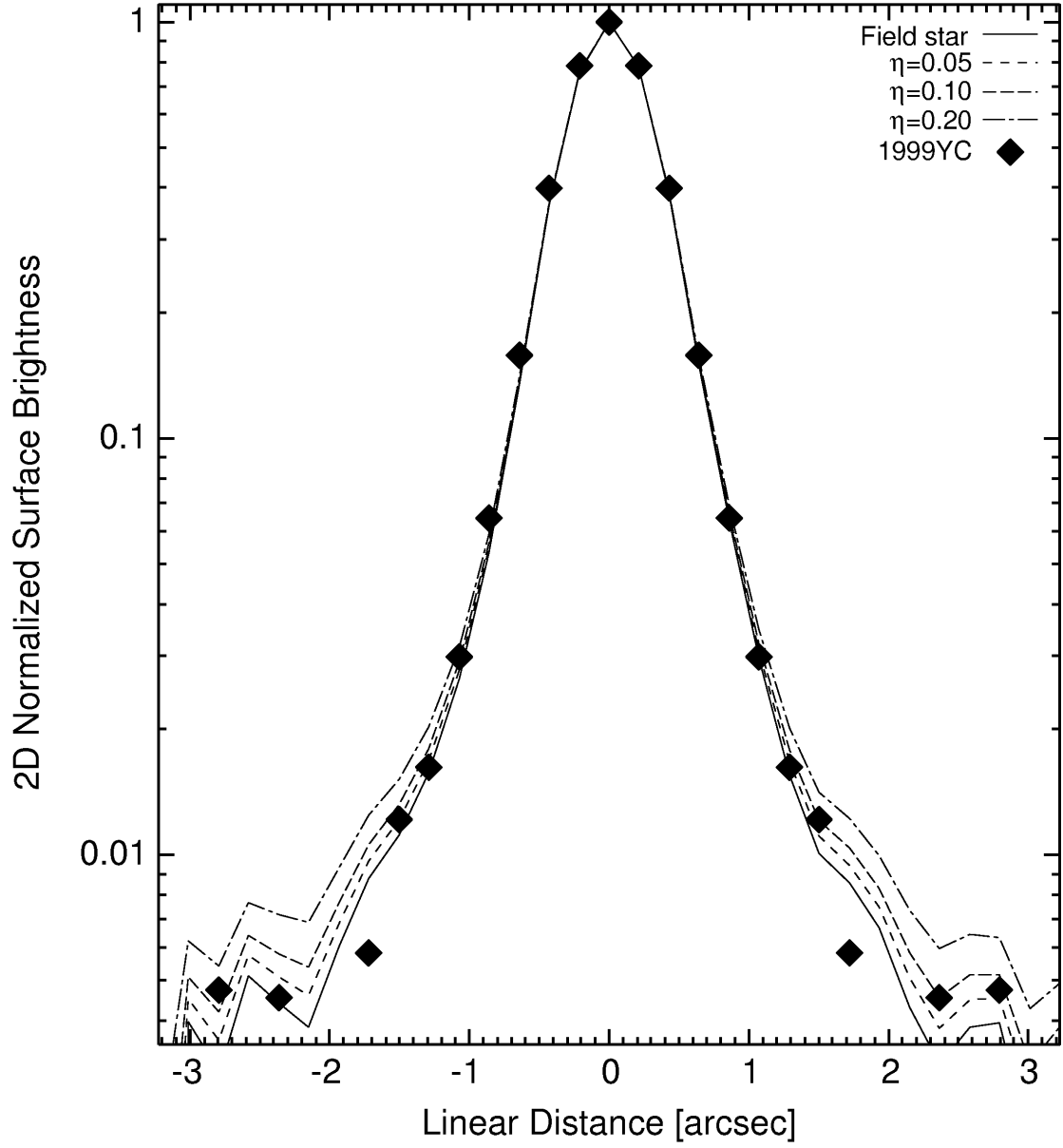}\caption{2D Surface brightness
models that compare 1999 YC's profile with seeing convolved models
having $\eta$=0.05, 0.10 and 0.20. (See Section 3.3) \label{2D}}
\end{figure*}

Figure \ref{2D} shows convolution models with coma levels of
$\eta$=0.05, 0.10 and 0.20, from which we estimate an upper limit
$\eta_{\rm lim}$ $\lesssim$0.1.  Assuming that the mass loss was
on-going and isotropic, the rate can be expressed as a function of the
parameter $\eta$ ~\citep{Luu1992}
\begin{equation}
\dot{M}=\frac{dM}{dt} = \frac{1.0 \times 10^{-3} 
\pi \rho \bar{a} \eta_{lim}r_{\rm obj}^2}{\phi
 R^{\frac{1}{2}}\Delta}
\end{equation}
where $\rho$=1000 ${\rm kg m^{-3}}$ is the assumed grain density,
$\bar{a}$=0.5$\times$$10^{-6}$\,m is the assumed grain radius, $r_{\rm
obj}$=700 m is 1999 YC's radius, and $R$ and ${\it \Delta}$ are given in
Table \ref{log}.  The mass loss rate was calculated $\dot{M}$
$\lesssim$2.4$\times$$10^{-3}$$\,{\rm kg\,s^{-1}}$ with $\eta_{\rm
lim}$$\lesssim$0.1.

It is not likely that water ice survives on the surface of 1999 YC or
any body with similarly small perihelion distance.  Nevertheless, it is
interesting to compute the maximum allowable fraction of the surface
that could be occupied by water ice while remaining consistent with the
point-like surface brightness profile of the object.

To do this, we convert $\dot{M}$ into the fraction of active area on the
surface of the object, {\it f}, via

\begin{equation}
f = \frac{\dot{M}}{4 \pi r^2_{\rm obj} \mu dm/dt}
\label{fraction}
\end{equation}

where $\mu$=1 is the assumed dust-to gas ratio.  Under the assumption
that volatile material (= water ice) exists, the specific sublimation
mass loss rate of water, $\rm dm/dt$ in ${\rm kg\,m^{-2}\,s^{-1}}$, is
calculated from the heat balance equation,

\begin{equation}
\frac{F_\odot (1-A)}{R^2} = \chi [\epsilon \sigma T^4 + L(T) dm/dt].
\end{equation}

Here, $F_\odot$ = 1365 W ${\rm m^{-2}}$ is the solar constant, R (in AU)
is the heliocentric distance, A = 0.11 is the assumed bond albedo
\citep{Green1985}, $\epsilon$ = 0.9 is the assumed emissivity, $\sigma$
= 5.67 $\times$ $10^{-8}$ W ${\rm m^{-2}}$ ${\rm K^{-4}}$ is the
Stephan-Boltzmann constant and $T$ in K is the equilibrium temperature.
The latent heat of sublimation for water $L(T)$ = (2.875 $\times$
$10^6$) -- (1.111 $\times$ $10^{3}$)$T$ in J ${\rm kg^{-1}}$ is taken
from the fit to $L(t)$ in \cite{Delsemme1971} .  The parameter $1\le
\chi \le4$ represents the distribution of solar energy over the surface
of object, where $\chi$ =1 corresponds to a flat slab facing the Sun,
$\chi$ = 2 to the standard thermal model (slow rotator) and $\chi$ =4 to
an isothermal sphere.  The term on the left represents the flux of
energy absorbed from the sun.  The terms on the right represent energy
lost from the nucleus surface by radiation and by latent heat of
sublimation.  In this first order calculation, thermal conduction is
neglected.

The specific sublimation mass loss rate can be derived iteratively using
the temperature dependent water vapor pressure given by
\cite{Fanale1984}.  At 2.6\,AU, assuming a flat slab model ($\chi$=1.0),
the maximum specific sublimation mass loss rate is $dm/dt$ = 4.3 $\times
10^{-5}$ ${\rm kg \,m^{-2}\,s^{-1}}$ and the temperature is 190\,K.  On
the other hand, minimum values of 9.7 $\times 10^{-7}$ $\rm
kg\,m^{-2}\,s^{-1}$ and 170 K are found using the isothermal model
($\chi$ =4), giving the maximum fraction of active area ${\it f}$ $\sim$
4.0 $\times$ $10^{-4}$ using eq.~(\ref{fraction}).

\begin{figure}[htbp]
\epsscale{0.9} \plotone{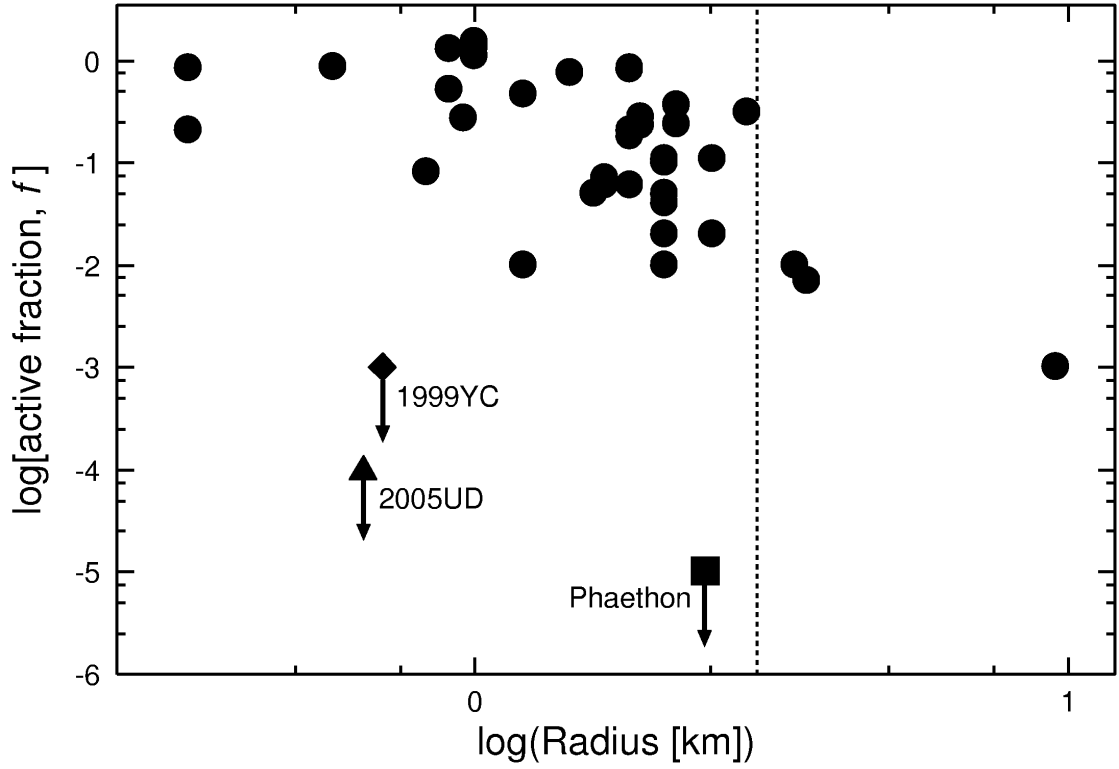} \caption{Radius vs. fractional active
area, ${\it f}$.  For JFCs with $r_{\rm obj}$ $\le$ 3\,km (vertical
dashed line), ${\it f}$ is an order of magnitude larger than the maximum
active fraction limit found in 1999 YC (see Sec.\ref{SBS}).
\label{frac}}
\end{figure}

Figure~\ref{frac} represents radius (km) versus fractional active area
${\it f}$ for 1999 YC, 2005 UD, 3200 Phaethon with determinations of $f$
for 27 Jupiter family comets (JFCs) \citep{Tancredi2006}.  The small
active surface fractions of the PGC candidates are obvious, with upper
limits of ${\it f}$ $<$ $10^{-3}$ on 1999 YC, on 2005 UD with ${\it f}$
$<$ $10^{-4}$ \citep{jewitt2006} and on 3200 Phaethon with ${\it f}$ $<$
$10^{-5}$ \citep{Hsieh2005}.  Relatively small active fractions are
found in 28P/Neujmin (${\it f}$= 0.001) and in 49P/Arend-Rigaux (${\it
f}$= 0.007), although these bodies are an order of magnitude larger than
1999 YC.  For bodies of comparable size, the upper limits to $f$ on PGC
are still smaller by more than one order of magnitude than on JFCs.

\section{Discussion}
The tiny limiting mass loss rates derived from observations of 1999 YC,
2005 UD and 3200 Phaethon (Table 6) can be compared with the total mass
of the Geminid meteoroid stream.  The stream has an age estimated
dynamically to be a few thousand years at the most (Jones 1978; Fox et
al. 1982; Jones \& Hawkes 1986; Gustafson 1989; Williams \& Wu 1993;
Ryabova 2001, see also summary in Jenniskens 2006) and the total mass is
$\sim$ $10^{12}$ -- $10^{13}$\,kg \citep{Hughes1989, jenniskens1994}.
Steady mass loss at the maximum rates allowed by the optical data,
namely $10^{-2}$ kg ${\rm s^{-1}}$ (see Table 6), would deliver only
$\sim$ 3$\times10^8$ kg in 1000 yrs.  Therefore, the large mass of the
Geminid meteoroid stream and the small allowable values of the current
mass production rates together point to origin of the stream by the
catastrophic break-up of the parent body, not by steady disintegration
at the observed rate \citep{jewitt2006} (see also Jenniskens 2008).

However, the mechanism responsible for the break-up of the PGC parent
body remains unknown.  \cite{jewitt2006} speculated that ice sublimation
in the core of the PGC parent body could be responsible for its
disintegration if the sublimation gas pressure substantially exceeded
the hydrostatic pressure (see also Samarasinha 2001).  This is possible
because a) the timescale for heat to conduct from the surface to the
core is smaller than the expected dynamical lifetime provided the radius
is $r \le$ 7 km and b) the orbitally-averaged temperature of a body in a
PGC-like orbit is high enough to promote strong sublimation of water ice
even in the core.  The main belt comets orbit in the asteroid belt (they
have asteroid T$_J >$ 3) and contain ice \citep{Hsieh2006}.  If such an
object were deflected into a planet crossing orbit with a small
perihelion distance, like the PGC bodies, it is conceivable that a
period of strong surface outgassing might be followed, after a thermal
diffusion time, by disruption due to sublimation of ice in the core.
Another possibility is that spin-up caused by torques from non-central
mass loss in such an object might result in centripetal disruption and
break-up, although whether this would produce a Geminid-like stream as
opposed to a few large chunks is far from clear.  Still another
possibility is that spin-up and disruption occur through the YORP
effect: the timescale for the action of YORP is short for bodies, like
1999 YC and the other members of the PGC, having small perihelion
distance and large eccentricities \citep{Scheeres2007}.

Measurements of the Na content of the Geminid meteoroids provide an
indicator of the effects of solar heating.  This is because Na is a
relatively volatile (temperature-sensitive) and abundant (easy to
detect) element.  Spectroscopic observations of Geminid meteors show an
extreme diversity of Na contents, from strong depletion of Na abundance
in some (ex. $\sim$ 7\% of the solar abundance) \citep{kasuga2005} to
sun-like values in others (Harvey 1973).  Line intensity ratios in the
Geminids also show a wide range of Na content, from undetectable to
strong \citep{Jiri2005}.  \citet{kasuga2006} investigated the thermal
desorption of Na in meteoroids in meteor streams during their orbital
motion in interplanetary space.  They found it unlikely that the Na
content has been modified thermally because the peak temperatures of the
meteoroids, even when at $q\sim 0.14$\,AU, are lower than the
sublimation temperature of alkali silicates ($\sim 900$\,K)
\citep{kasuga2006}.  Therefore, the diversity of Na abundances observed
in Geminid meteoroids must have another origin, perhaps related to the
thermal evolution of 3200 Phaethon or the larger sized fragments
themselves.  For example, the Na content may relate to the position in
the parent body before the meteoroids were ejected.  The physical
properties of meteoroids from the surface regions could be changed by
compaction associated with loss of volatiles \citep{beech1984}.  Those
Geminid meteoroids would be stronger and have higher bulk density
\citep{verniani1967,Wetherill1986,baba2002}.  On the other hand,
Geminids from the interior might be relatively fresh uncompacted and
volatile rich, with the Na preserving more Sun-like values.  Eventually,
the true natures of PGC-fragments and ice-rich asteroids (dormant
comets) may be revealed by missions resembling NASA's ``Deep Impact''
\citep{AHearn2005,kasuga2006b}.

\section{Summary}

Optical observations of asteroid 1999 YC, a suggested member of the
Phaethon-Geminid Stream Complex, give the following results.
 	
\begin{enumerate}
\item Optical colors measured for 1999 YC are nearly neutral, consistent
      with those of the taxonomic C-type asteroids and slightly redder
      than the neutral-blue colors found on other Phaethon-Geminid
      stream complex bodies.
\item The absolute red magnitude is ${\rm m}_{\rm
      R}$(1,1,0)=16.96$\pm$0.03, giving the equivalent circular diameter
      $D_{\rm e}$=1.4$\pm$0.1\,km assuming the same geometric albedo as
      3200 Phaethon ($p_R$=0.11).
\item The light curve of 1999 YC has the double peaked period of $P_{\rm
      rot}$=4.4950 $\pm$ 0.0010\,hr.  The photometric range of $\Delta
      m_R$=0.69$\pm$0.05 mag corresponds to an axis ratio of
      1.89$\pm$0.09, suggesting an elongated body with the critical
      density $\ge$ 1000\,kg\,${\rm m^{-3}}$.
\item No evidence of lasting mass loss was found from the surface
      brightness profiles in imaging data.  The maximum mass loss rate
      is $\sim$$10^{-3}$$\,{\rm kg\,s^{-1}}$ which corresponds to the
      fractional active area ${\it f}$ $<$ $10^{-3}$.
\item Catastrophic breakup or comet-like disintegration of a precursor
      body are suggested because the mass loss rates are too small to
      form the massive Geminid meteoroid stream in steady state given
      the 1000 yr dynamical lifetime.
\end{enumerate}

\acknowledgments

 T.K is grateful to Robert Jedicke for his enthusiastic encouragement in
 studies at the Institute for Astronomy, UH.  We thank our colleagues
 Bin Yang, Pedro Lacerda (UH), Katsuhito Ohtsuka (Tokyo Meteor Network),
 Masanao Abe, Sunao Hasegawa, Kyoko Kawakami, Taichi Kawamura
 (ISAS/JAXA), and Daisuke Kinoshita (National Central Univ., Taiwan) for
 useful discussions.  We appreciate financial support to T.K. from the
 JSPS Research Fellowships for young scientists and to D.J. from NASA's
 Planetary Astronomy Program.  Lastly, we thank Peter Jenniskens for his
 prompt review.

\clearpage

\begin{deluxetable}{llccccccccc}
\tabletypesize{\scriptsize}
\tablecaption{Observation Log \label{log}}
\tablewidth{0pt}
\tablehead{
 &   &  & Integration &  &  R\tablenotemark{b}  &
 $\Delta$\tablenotemark{c} & $\alpha$ \tablenotemark{d} \\
Object   &UT Date & Telescope \tablenotemark{a}&  (s) & Filter & (AU) &
 (AU) & (deg) \\
}
\startdata
1999YC&2007 Oct. 4  & UH\,2.2  & 400    &  1\,B, 1\,V, 13\,R  & 2.6030 &
 1.9121 & 18.67\\
&2007 Oct. 12 & Keck\,10 & 200    &  2$\times$(B \& R), 2$\times$(B \& V)     & 2.6013 & 1.8185 & 16.37\\
&2007 Oct. 18 & UH\,2.2  & 300&  48\,R & 2.5985 & 1.7592 & 14.43\\
 
 &2007 Oct. 19 & UH\,2.2  & 300 &  13\,R & 2.5979 & 1.7499 & 14.08\\
155140 (2005\,UD)&2007 Oct. 4 & UH\,2.2  & 300    &  2\,B, 2\,V, 2\,R & 2.3767 & 1.6423 & 19.81\\ 
(3200) Phaethon&2007 Sep. 4 & UH\,2.2  & 200    &  1\,B, 1\,R, 1\,I & 1.9916 & 2.0799 & 28.58 \\
& &   & 100    &  2\,B, 3\,V, 2\,R, 2\,I &     &   & \\
\enddata
\tablenotetext{a}{UH\,2.2 = University of Hawaii 2.2\,m (88-inch)
 telescope, Keck\,10 = 10\,m Keck-I telescope}
 \tablenotetext{b}{Heliocentric distance} \tablenotetext{c}{Geocentric
 distance} \tablenotetext{d}{Phase angle}
\end{deluxetable}
\clearpage


\begin{deluxetable}{lcccccccc}
\tabletypesize{\scriptsize}
\tablewidth{0pt}
\tablecaption{Color photometry for 1999YC \label{colphot}}
\tablehead{
\colhead{Object} & \colhead{Telescope$^{a}$} &
\colhead{Date (UT\,2007)}      &
\colhead{Midtime}          &
\colhead{B - R}  & \colhead{B - V} & \colhead{V - R} &\colhead{R} }         
\startdata
1999YC& UH\,2.2   & Oct.4&  11.09992 &1.07$\pm$0.07&-&-              & 21.15$^{b}$    \\
          	  &&     &  11.34678 & -           &-& 0.40 $\pm$0.06& 21.13$^{b}$    \\
         &Keck10& Oct.12 & 205.20989 &1.110$\pm$0.013 &- &-          & 21.120$\pm$0.008\\
      & 	&	 & 205.40086 &  -    &0.734$\pm$0.011&-      & 21.00$^{b}$\\
      &         &        & 205.47228 &  -    &0.728$\pm$0.011 &-     & 20.96$^{b}$\\
      &         &        & 205.55449 &1.112$\pm$0.011&-&-            & 20.911$\pm$0.008\\
\enddata
\tablenotetext{a}{UH\,2.2 = University of Hawaii 2.2\,m telescope,
 Keck\,10 = 10\,m Keck-I telescope}
\tablenotetext{b}{$R$-band magnitude interpolated from the light curve
 in Fig.\ref{Abs} on UH\,2.2 and from the linear fitting on Keck\,10.}
\end{deluxetable}

\clearpage


\begin{deluxetable}{lccccc}
\tabletypesize{\scriptsize}
\tablewidth{0pt}
\tablecaption{Light curve photometry through {\it R}-band filter on
 UH\,2.2 \label{lightphot}}
\tablehead{
\colhead{N} & \colhead{Date (UT 2007)}      &
\colhead{Midtime\tablenotemark{a}}& \colhead{Apparent:$m_{\rm R}$\tablenotemark{b}}  &
\colhead{Relative\tablenotemark{c}} }         
\startdata
1  & Oct.4 & 10.60783 & 21.527$\pm$0.052&   0.002$\pm$ 0.052& \\
2  & Oct.4 & 10.73076 & 21.393$\pm$0.048&  -0.160$\pm$ 0.048& \\
3  & Oct.4 & 10.85411 & 21.260$\pm$0.040&  -0.269$\pm$ 0.040& \\
4  & Oct.4 & 10.97708 & 21.198$\pm$0.035&  -0.315$\pm$ 0.035& \\
5  & Oct.4 & 11.48112 & 21.102$\pm$0.037&  -0.395$\pm$ 0.037& \\
6  & Oct.4 & 11.60382 & 21.282$\pm$0.045&  -0.220$\pm$ 0.045& \\
7  & Oct.4 & 11.72611 & 21.270$\pm$0.038&  -0.232$\pm$ 0.038& \\
8  & Oct.4 & 11.84932 & 21.378$\pm$0.042&  -0.131$\pm$ 0.042& \\
9  & Oct.4 & 11.97159 & 21.537$\pm$0.053&   0.030$\pm$ 0.053& \\
10 & Oct.4 & 12.09383 & 21.771$\pm$0.080&   0.226$\pm$ 0.080& \\
11 & Oct.4 & 12.21696 & 22.012$\pm$0.112&   0.350$\pm$ 0.112& \\
12 & Oct.4 & 12.33933 & 21.788$\pm$0.075&   0.254$\pm$ 0.075& \\
13 & Oct.4 & 12.46161 & 21.844$\pm$0.070&   0.301$\pm$ 0.070& \\
14 & Oct.18&346.33945 & 20.791$\pm$0.033&  -0.281$\pm$ 0.033& \\
15 & Oct.18&346.43490 & 20.699$\pm$0.024&  -0.320$\pm$ 0.024& \\
16 & Oct.18&346.53053 & 20.853$\pm$0.026&  -0.211$\pm$ 0.026& \\
17 & Oct.18&346.62621 & 20.937$\pm$0.023&  -0.190$\pm$ 0.023& \\
18 & Oct.18&346.72193 & 20.854$\pm$0.024&  -0.164$\pm$ 0.024& \\
19 & Oct.18&346.81773 & 20.877$\pm$0.029&  -0.122$\pm$ 0.029& \\
20 & Oct.18&346.91605 & 21.016$\pm$0.028&   0.005$\pm$ 0.028& \\
21 & Oct.18&347.01132 & 21.124$\pm$0.029&   0.077$\pm$ 0.029& \\
22 & Oct.18&347.10691 & 21.302$\pm$0.039&   0.209$\pm$ 0.039& \\
23 & Oct.18&347.20252 & 21.305$\pm$0.042&   0.302$\pm$ 0.042& \\
24 & Oct.18&347.29782 & 21.330$\pm$0.042&   0.293$\pm$ 0.042& \\
25 & Oct.18&347.39352 & 21.356$\pm$0.044&   0.311$\pm$ 0.044& \\
26 & Oct.18&347.48865 & 21.326$\pm$0.037&   0.282$\pm$ 0.037& \\
27 & Oct.18&347.58421 & 21.216$\pm$0.033&   0.209$\pm$ 0.033& \\
28 & Oct.18&347.67977 & 21.097$\pm$0.032&   0.071$\pm$ 0.032& \\
29 & Oct.18&347.77528 & 21.048$\pm$0.027&   0.035$\pm$ 0.027& \\
30 & Oct.18&347.87110 & 20.951$\pm$0.026&  -0.041$\pm$ 0.026& \\
31 & Oct.18&347.96628 & 20.636$\pm$0.024&  -0.361$\pm$ 0.024& \\
32 & Oct.18&348.06137 & 20.858$\pm$0.032&  -0.139$\pm$ 0.032& \\
33 & Oct.18&348.15678 & 20.764$\pm$0.032&  -0.237$\pm$ 0.032& \\
34 & Oct.18&348.25232 & 20.786$\pm$0.030&  -0.200$\pm$ 0.030& \\
35 & Oct.18&348.34741 & 20.731$\pm$0.027&  -0.233$\pm$ 0.027& \\
36 & Oct.18&348.44813 & 20.662$\pm$0.036&  -0.322$\pm$ 0.036& \\
37 & Oct.18&348.54359 & 20.788$\pm$0.044&  -0.195$\pm$ 0.044& \\
38 & Oct.18&349.02222 & 20.737$\pm$0.070&  -0.255$\pm$ 0.070& \\
39 & Oct.18&349.11891 & 20.852$\pm$0.055&  -0.138$\pm$ 0.055& \\
40 & Oct.18&349.21456 & 20.852$\pm$0.082&  -0.139$\pm$ 0.082& \\
41 & Oct.18&349.32338 & 21.210$\pm$0.044&   0.224$\pm$ 0.044& \\
42 & Oct.18&349.41889 & 21.264$\pm$0.048&   0.281$\pm$ 0.048& \\
43 & Oct.18&349.51430 & 21.282$\pm$0.063&   0.298$\pm$ 0.063& \\
44 & Oct.18&349.60946 & 21.356$\pm$0.077&   0.373$\pm$ 0.077& \\
45 & Oct.18&349.70458 & 21.340$\pm$0.052&   0.358$\pm$ 0.052& \\
46 & Oct.18&349.79988 & 21.329$\pm$0.054&   0.348$\pm$ 0.054& \\
47 & Oct.18&349.89582 & 21.186$\pm$0.085&   0.207$\pm$ 0.085& \\
48 & Oct.18&349.99088 & 21.031$\pm$0.047&   0.050$\pm$ 0.047& \\
49 & Oct.18&350.10853 & 20.992$\pm$0.029&   0.033$\pm$ 0.029& \\
50 & Oct.18&350.20475 & 20.891$\pm$0.027&  -0.072$\pm$ 0.027& \\
51 & Oct.18&350.30036 & 20.742$\pm$0.023&  -0.241$\pm$ 0.023& \\
52 & Oct.18&350.39558 & 20.784$\pm$0.024&  -0.198$\pm$ 0.024& \\
53 & Oct.18&350.49086 & 20.754$\pm$0.022&  -0.229$\pm$ 0.022& \\
54 & Oct.18&350.58617 & 20.731$\pm$0.022&  -0.249$\pm$ 0.022& \\
55 & Oct.18&350.68131 & 20.725$\pm$0.020&  -0.256$\pm$ 0.020& \\
56 & Oct.18&350.77684 & 20.734$\pm$0.021&  -0.250$\pm$ 0.021& \\
57 & Oct.18&350.87236 & 20.699$\pm$0.022&  -0.282$\pm$ 0.022& \\
58 & Oct.18&350.96777 & 20.722$\pm$0.021&  -0.259$\pm$ 0.021& \\
59 & Oct.18&351.06372 & 20.773$\pm$0.025&  -0.204$\pm$ 0.025& \\
60 & Oct.18&351.16056 & 20.772$\pm$0.024&  -0.210$\pm$ 0.024& \\
61 & Oct.18&351.25728 & 20.892$\pm$0.029&  -0.088$\pm$ 0.029& \\
62 & Oct.19&370.92062 & 20.691$\pm$0.026&  -0.296$\pm$ 0.026& \\
63 & Oct.19&371.01599 & 20.731$\pm$0.025&  -0.264$\pm$ 0.025& \\
64 & Oct.19&371.11155 & 20.845$\pm$0.026&  -0.225$\pm$ 0.026& \\
65 & Oct.19&371.20698 & 20.842$\pm$0.026&  -0.222$\pm$ 0.026& \\
66 & Oct.19&371.30237 & 20.794$\pm$0.025&  -0.242$\pm$ 0.025& \\
67 & Oct.19&371.39801 & 20.852$\pm$0.029&  -0.219$\pm$ 0.029& \\
68 & Oct.19&371.49356 & 20.863$\pm$0.022&  -0.119$\pm$ 0.022& \\
69 & Oct.19&371.58887 & 20.880$\pm$0.031&  -0.139$\pm$ 0.031& \\
70 & Oct.19&371.68409 & 20.897$\pm$0.028&  -0.071$\pm$ 0.028& \\
71 & Oct.19&371.77922 & 20.994$\pm$0.034&  -0.005$\pm$ 0.034& \\
72 & Oct.19&371.87735 & 20.967$\pm$0.038&   0.019$\pm$ 0.038& \\
73 & Oct.19&371.97288 & 21.184$\pm$0.037&   0.235$\pm$ 0.037& \\
74 & Oct.19&372.07014 & 21.273$\pm$0.040&   0.322$\pm$ 0.040& \\
\enddata
\tablenotetext{a}{Time since UT 2007 October 4.00000.  The middle of
 integration times is taken.}
\tablenotetext{b}{Apparent magnitude measured in R-band image.}
\tablenotetext{c}{Relative red magnitude to 7 field stars in background.}
\end{deluxetable}


\clearpage


\begin{deluxetable}{lcccl}
\tabletypesize{\scriptsize}
\tablewidth{0pt}
\tablecaption{Color results of Phaethon-Geminid Complex \label{color}}
\tablehead{
\colhead{Object} & \colhead{B - V}      &
\colhead{V - R}          & \colhead{R - I}         &
\colhead{Source} }         
\startdata
1999YC  & 0.71$\pm$0.04 &0.36$\pm$0.03& --          &(1) \\
2005UD  & 0.68$\pm$0.01 &0.39$\pm$0.02  & --          &(1)\\
        & 0.66$\pm$0.03 &0.35$\pm$0.02  &0.33$\pm$0.02&(2)\\
        & 0.63$\pm$0.01 &0.34$\pm$0.01  &0.30$\pm$0.01&(3)\\
Phaethon& 0.61$\pm$0.01 &0.34$\pm$0.03  &0.27$\pm$0.04&(1) \\
        & 0.59$\pm$0.01 &0.35$\pm$0.01  &0.32$\pm$0.01&(4)\\
\smallskip
        & --    & 0.34  & --                          &(5)\\
Solar colors & 0.67  &0.36  & 0.35  &(2)\\
\enddata
\tablerefs{(1) This work; (2) \cite{jewitt2006}; (3) \cite{Kinoshita2007} ; (4)
 \cite{Dundon2005} ; (5) \cite{Skiff1996}}
\end{deluxetable}

\clearpage

\begin{deluxetable}{cccc}
\tabletypesize{\scriptsize}
\tablecaption{Absolute Red magnitude ${\rm m}_{\rm R}$(1,1,0) and 
Equivalent circular $D_{\rm e}$ \label{size}}
\tablewidth{0pt}
\tablehead{
Object & ${\rm m}_{\rm R}$(1,1,0) & $D_{\rm e}$(km) & Source}
\startdata
1999YC  & 16.96$\pm$0.03 & 1.4$\pm$0.1 & (1) \\
2005UD  & 17.23$\pm$0.03 & 1.2$\pm$0.1 & (1) \\
        & 17.13$\pm$0.03 & 1.3$\pm$0.1 & (2)\\
Phaethon& 14.22$\pm$0.01&4.9$\pm$0.4  &(1) \\
        & $\sim$14.3$\pm$0.1 &4.7$\pm$0.5 & (3),(4)\\
\enddata
\tablerefs{(1) This work; (2) \cite{jewitt2006}; (3) \cite{Green1985} ; (4)
 \cite{Dundon2005}}
\end{deluxetable} 
\clearpage

\begin{deluxetable}{lcccc}
\tabletypesize{\scriptsize}
\tablecaption{Physical properties of 1999YC, 2005\,UD and 3200 Phaethon\label{results}}
\tablewidth{0pt}
\tablehead{
Quantity & Symbol & 1999YC\,$^{a}$ & 2005\,UD\,$^{b}$ & 3200 Phaethon\,$^{c,d}$}
\startdata
Semimajor axis         & $a$  & 1.422 & 1.275 &1.271  \\
Perihelion             & $q$  & 0.241 & 0.163 &0.140  \\
Eccentricity           & $e$  & 0.831 & 0.872 &0.890  \\
Inclination            & $i$  & 38.16 & 28.75 &22.16  \\
Rotational period (hr)  & $P_{\rm rot}$&4.495 & 5.249 & 3.59\\
Photometric range (mag)& $m_{\rm R}$  &0.69$\pm$0.05 & 0.40$\pm$0.05	& 0.4\\
Critical density (${\rm kg\,m^{-3}}$) & $\rho_c$&1000  & 570 &1200 \\
Mass loss rate (${\rm kg\,s^{-1}}$) & $\dot{M}$  & 0.001 & 0.01 &0.01  \\
Fractional active area & ${\it f}$& $<$ $10^{-3}$ & $<$ $10^{-4}$ &$<$ $10^{-5}$ \\
\enddata
\tablecomments{Orbital data are from \cite{Ohtsuka2006} and NASA JPL HORIZON.}
\tablenotetext{a}{This work}
\tablenotetext{b}{\cite{jewitt2006}}
\tablenotetext{c}{\cite{Dundon2005}}
\tablenotetext{d}{\cite{Hsieh2005}} 
\end{deluxetable} 

\clearpage

\clearpage


\clearpage


\clearpage

\clearpage

\clearpage

\clearpage

\clearpage


\clearpage
\newpage







\clearpage

\end{document}